\documentclass[12pt]{article}
\usepackage{latexsym}\usepackage{graphicx}\usepackage{epstopdf}
\usepackage{amsmath, amssymb}\usepackage{amsthm}

\newcommand{\bfx}{\mbox{\boldmath $x$}}
\newcommand{\bfy}{\mbox{\boldmath $y$}}\newcommand{\bfv}{\mbox{\boldmath $v$}}

\newcommand{\bfnu}{\mbox{\boldmath $\nu$}}\newcommand{\bfr}{\mbox{\boldmath $r$}}

\newcommand{\beq}{\begin{equation}}\newcommand{\eeq}{\end{equation}}
\newtheorem{theorem}{Theorem}
\newtheorem{corollary}{Corollary}

\newtheorem{lemma}{Lemma}
\def\beginofproof{\begin{proof}}
\newtheorem{exam}{{\bf Example}}
\def\endofproof{\end{proof}}
\usepackage{cite} 
\DeclareMathOperator*{\argmin}{arg\,min}
\def\mysingleq#1{``#1''}
\newcommand{\bZ}{\mathbb{Z}}
\newcommand{\bR}{\mathbb{R}}
\newcommand{\bc}{\mathbf{c}}
\newcommand{\GCD}{\mathrm{GCD}}
\begin{document}
\title{Pearson Codes
}
\author{Jos H. Weber\thanks{Jos H. Weber is with Delft University of Technology, Delft, The Netherlands. E-mail: j.h.weber@tudelft.nl.}\\
Kees A. Schouhamer Immink\thanks{Kees A. Schouhamer Immink is with Turing Machines Inc, Willemskade 15b-d, 3016 DK Rotterdam, The Netherlands. E-mail: immink@turing-machines.com.}\\
and\\
Simon  R. Blackburn\thanks{Simon R. Blackburn is with the Department of Mathematics, Royal Holloway University of London, Egham, Surrey TW20 0EX, United Kingdom. E-mail: S.Blackburn@rhul.ac.uk}}
\maketitle

\begin{abstract}
The Pearson distance has been advocated for improving the error performance of noisy channels with unknown gain and offset. The Pearson distance can only fruitfully be used for sets of $q$-ary codewords, called {\it Pearson codes}, that satisfy specific properties. We will analyze constructions and properties of optimal Pearson codes.
We will compare the redundancy of optimal Pearson codes with the redundancy of prior art $T$-constrained codes, which consist of $q$-ary sequences in which $T$ pre-determined reference symbols appear at least once. In particular, it will be shown that for $q\le 3$ the $2$-constrained codes are optimal Pearson codes, while for $q\ge 4$ these codes are not optimal.

{\bf Key words:} flash memory, digital optical recording, Non-Volatile Memory, NVM,  Pearson distance.
\end{abstract}

\section{Introduction}

In non-volatile memories, such as floating gate memories, the data is represented by stored charge, which can leak away from the floating gate. This leakage may result in a shift of the offset or {\it threshold voltage} of the memory cell. The amount of leakage depends on the time elapsed between writing and reading the data. As a result, the offset between different groups of cells may be very different so that prior art automatic offset or gain control, which estimates the mismatch from the previously received data, can not be applied. Methods to solve these difficulties in Flash memories have been discussed in, for example,~\cite{Jia}, \cite{Sa3}, \cite{Zh1}, \cite{Sa5}. In optical disc media, such as the popular Compact Disc, DVD, and Blu-ray disc, the retrieved signal depends on the dimensions of the written features and upon the quality of the light path, which may be obscured by fingerprints or scratches on the substrate. Fingerprints and scratches will result in rapidly varying offset and gain variations of the retrieved signal. Automatic gain and offset control in combination with dc-balanced codes are applied albeit at the cost of redundancy~\cite{I59}, and thus improvements to the art are welcome.

Immink \& Weber~\cite{I67} showed that detectors that use the Pearson distance offer immunity to offset and gain mismatch. The Pearson distance can only be used for a set of codewords with special properties, called a {\it Pearson set} or {\it Pearson code}. Let ${\cal S}$ be a codebook of chosen $q$-ary codewords $\bfx=(x_1, x_2,\ldots, x_n)$ over the $q$-ary alphabet ${\cal Q}=\{0,1, \ldots, q-1\}$, $q \geq 2$, where $n$, the {\it length} of $\bfx$, is a positive integer. Note that the alphabet symbols are to be treated as being just integers rather than  elements of $\mathbb{Z}_q$. A Pearson code with maximum possible size given the parameters $q$ and $n$ is said to be {\it optimal}.

In Section~\ref{detect}, we set the stage with a description of Pearson distance detection and the properties of the constrained codes used in conjunction with it. Section~\ref{description} gives a description of $T$-constrained codes, a type of code described in the prior art~\cite{I67}, used in conjunction with the Pearson distance detector, while Section~\ref{gencon} offers a general construction of optimal Pearson codes and a computation of their cardinalities. The rates of $T$-constrained codes will be compared with optimal rates of Pearson codes. In Section~\ref{conclusions}, we will describe our conclusions.
\section{Preliminaries}\label{detect}

We use the shorthand notation $a\bfv+b$ = $(av_1+b, av_2+b, \ldots, av_n+b)$. In~\cite{I67}, the authors suppose a situation where the sent codeword, $\bfx$, is received as the vector $\bfr = a(\bfx+\bfnu)+b$, $r_i \in \mathbb{R}$. Here $a$ and $b$ are unknown real numbers with $a$ positive, called the \emph{gain} and the (dc-) \emph{offset} respectively. Moreover, $\bfnu$ is an additive noise vector: $\bfnu$ = $(\nu_1, \ldots, \nu_n)$,  where $\nu_i \in \mathbb{R}$ are noise samples from a zero-mean Gaussian distribution. Note that both gain and offset do not vary from symbol to symbol, but are the same for all $n$ symbols.


The receiver's ignorance of the channel's momentary gain and offset may lead to massive performance degradation as shown, for example, in~\cite{I67} when a traditional detector, such as threshold or maximum likelihood detector, is used. In the prior art, various methods have been proposed to overcome this difficulty. In a first method, data reference, or `training', patterns are multiplexed with the user data in order to `teach' the data detection circuitry the momentary values of the channel's characteristics such as impulse response, gain, and offset. In a channel with unknown gain and offset, we may use two reference symbol values, where in each codeword, a first symbol is set equal to the lowest signal level and a second symbol equal to the highest signal level. The positions and amplitudes of the two reference symbols are known to the receiver. The receiver can straightforwardly measure the amplitude of the retrieved reference symbols, and normalize the amplitudes of the remaining symbols of the retrieved codeword before applying detection. Clearly, the redundancy of the method is two symbols per codeword.

In a second prior art method, codes satisfying equal balance and energy constraints\cite{I62}, which are immune to gain and offset mismatch, have been advocated. The redundancy of these codes, denoted by $r_0$, is given by~\cite{I62}
\beq  r_0 \approx \log_q n + \log_q (q^2-1)\sqrt{q^2-4} + \log_q \frac{\pi }{12\sqrt {15}} .\label{r_0} \eeq
In a recent contribution, Pearson distance detection is advocated since its redundancy is much less than that of balanced codes~\cite{I67}. The Pearson distance between the vectors $\bfx$ and $\hat\bfx$ is defined as follows. For a vector \bfx, define
\beq \overline{\bfx} =  \frac{1}{n} \sum_{i=1}^n  x_i , \label{eqms6b}\eeq
and
\beq \sigma^2_{\bfx} =  \sum_{i=1}^n ( x_i - \overline{\bfx})^2 \label{eqms6a}. \eeq
Note that $\sigma_{\bfx}$ is closely related to, but not the same as, the standard deviation of $\bfx$. 
The {\it (Pearson) correlation coefficient} is defined by
\beq \rho_{ \bfx, \hat\bfx} = \frac{ \sum_{i=1}^n (x_i - \overline {\bfx})(\hat x_i - \overline {\hat \bfx} )} {\sigma_{\bfx} \sigma_{\hat \bfx } }, \label{eqms6cc} \eeq
and the \emph{Pearson distance} is given by
\beq  \delta(\bfx, \hat{\bfx}) = 1 - \rho_{ \bfx, \hat{\bfx}}. \label{eqms6c} \eeq
The Pearson distance and Pearson correlation coefficient are well-known concepts in statistics and cluster analysis. Note that we have $|\rho_{\bfx, \hat{\bfx}}| \leq 1$ by a corollary of the Cauchy-Schwarz Inequality\cite[Section IV.4.6]{MGB74}, which implies that $0 \leq \delta(\bfx, \hat{\bfx})\leq 2$.

A minimum Pearson distance detector outputs the codeword
$$ {\bfx_o} = \argmin_{\hat{\bfx}\in {\cal S}} \delta (\bfr, \hat\bfx). $$
As the Pearson distance is translation and scale invariant, that is,
$$ \delta(\bfx, \hat{\bfx}) = \delta(a\bfx+b, \hat{\bfx}), $$
we conclude that the Pearson distance between the vectors $\bfx$ and $\hat\bfx$ is independent of the channel's gain or offset mismatch, so that, as a result, the error performance of the minimum Pearson distance detector is immune to gain and offset mismatch. This virtue implies, however, that the minimum Pearson distance detector cannot be used in conjunction with arbitrary codebooks, since
$$  \delta (\bfr, \hat\bfx) = \delta (\bfr, \hat\bfy)  $$
if $\hat\bfy = c_1\hat\bfx+c_2$, $c_1,c_2\in \mathbb{R}$ and $c_1>0$. In other words, since a minimum Pearson detector cannot distinguish between the words $\hat\bfx$ and $\hat\bfy=c_1\hat\bfx+c_2$, the codewords must be taken from a codebook ${\cal S}\subseteq{\cal Q}^n$ that  guarantees unambiguous detection with the Pearson distance metric (\ref{eqms6c}).

It is a well-known property of the Pearson correlation coefficient, $\rho_{\bfx, \hat{\bfx}}$, that
$$ \rho_{\bfx, \hat{\bfx}} = 1 $$
if and only if
$$ \hat\bfx = c_1+c_2 \bfx , $$
where the coefficients $c_1$ and $c_2>0$ are real numbers \cite[Section IV.4.6]{MGB74}. It is further immediate, see (\ref{eqms6cc}), that the Pearson distance is undefined for codewords $\bfx$ with $\sigma_{\bfx}=0$, i.e., for multiples of the all-one vector. We coined the name {\it Pearson code} for a set of codewords that can be uniquely decoded by a minimum Pearson distance detector. We conclude that codewords in a Pearson code must satisfy two conditions, namely
\begin{itemize}
\item {\it Property A:} If $\bfx \in {\cal S}$ then $c_1+c_2 \bfx \notin {\cal S}$ for all $c_1,c_2\in \mathbb{R}$ with $(c_1,c_2)\ne (0,1)$ and $c_2>0$.
\item {\it Property B:} $\bfx = (c,c,\ldots,c) \notin {\cal S}$ for all $c\in \mathbb{R}$ .
\end{itemize}
In the remaining part of this paper, we will study constructions and properties of Pearson codes. In particular, we are interested in Pearson codes that are optimal in the sense of having the largest number of codewords for given parameters $n$ and $q$. We will commence with a description of prior art $T$-constrained codes, a first example of Pearson codes.
\section{$T$-constrained codes}\label{description}

For integers $T$ satisfying $1\le T \leq q$, $T$-constrained codes~\cite{I66}, denoted by ${\cal S}_{q,n}(a_1,\ldots, a_T)$, consist of $q$-ary codewords of length $n$, where $T$ {\em preferred} or {\em reference} symbols $a_1,\ldots, a_T$ $\in {\cal Q}$ must each appear at least once in a codeword. Thus, each codeword, $(x_1, x_2, \ldots, x_n)$, in a $T$-constrained code satisfies
$$|\{i:x_i=j\}|>0 \mbox{ for each } j \in \{ a_1, \ldots, a_T\}.$$
The number of $q$-ary sequences of length $n$, $N_T(q,n)$, where $T$ distinct pre-defined symbols occur at least once in every sequence, equals~\cite{I66}
\beq  N_T(q,n) = \sum_{i=0}^T (-1)^i  \left( \substack { T \\ {T-i} }   \right) (q-i)^n, \,\, n \geq T. \label{eqnT} \eeq
For example, we easily find for $T=1$ and $T=2$ that
\beq N_1(q,n) =q^n-(q-1)^n \label{eqN1} \eeq
and
\beq N_2(q,n) = q^n-2(q-1)^n+(q-2)^n. \label{eqN2} \eeq
Clearly, the number of $T$-constrained sequences is not affected by the choice of the specific $T$ symbols we like to favor.

For the binary case, $q=2$, we simply find that ${\cal S}_{2,n}(0)$ is obtained by removing the all-`1' word from ${\cal Q}^n$, that ${\cal S}_{2,n}(1)$ is obtained by removing the all-`0' word from ${\cal Q}^n$, and that ${\cal S}_{2,n}(0,1)$ is obtained by removing both the all-`1' and all-`0' words from ${\cal Q}^n$, where ${\cal Q}=\{0,1\}$. Hence, indeed, $$N_1(2,n)=2^n-1$$ and $$N_2(2,n)=2^n-2.$$

The $2$-constrained code ${\cal S}_{q,n}(0,q-1)$ is a Pearson code as it satisfies Properties A and B~\cite{I67}. There are more examples of $2$-constrained sets that are Pearson codes, such as ${\cal S}_{q,n}(0,1)$. Note, however, that not all $2$-constrained sets are Pearson codes. For example, ${\cal S}_{q,n}(0,2)$ does not satisfy Property A if $q\ge 5$, since, e.g., both $(0,1,2,\ldots,2)$ and $(0,2,4,\ldots,4) = 2\times(0,1,2,\ldots,2)$ are codewords

It is obvious from Property B that the  code ${\cal S}_{2,n}(0,1)$ of size $2^n-2$ is the optimal binary Pearson code. For the ternary case, $q=3$, it can easily be argued that ${\cal S}_{3,n}(0,1)$, ${\cal S}_{3,n}(0,2)$, and ${\cal S}_{3,n}(1,2)$ are all optimal Pearson codes of size $3^n-2^{n+1}+1$.

However, for $q>3$ the 2-constrained sets such as ${\cal S}_{q,n}(0,1)$, ${\cal S}_{q,n}(0,q-1)$, and ${\cal S}_{q,n}(q-2,q-1)$, all of size $N_2(q,n)$, are not optimal Pearson codes, except when $n=2$. For example, for $q=4$, it can be easily checked that the set ${\cal S}_{4,n}(0,3) \cup{\cal S}_{3,n}(0,1,2)$ is a Pearson code. Its size equals $N_2(4,n)+N_3(3,n) = 4^n-3^n-2^{n+1}+3$, which is larger than $N_2(4,n)$ and actually turns out to be the maximum possible size of any Pearson code for $q=4$, as shown in the next section, where we will address the problem of constructing optimal Pearson codes for any value of $q$.

\section{Optimal Pearson codes}\label{gencon}
For $\bfx=(x_1,x_2,\ldots,x_n)\in{\cal Q}^n$, let $m(\bfx)$ and $M(\bfx)$ denote the smallest and largest value, respectively, among the $x_i$. Furthermore, in case $\bfx$ is not the all-zero word, let GCD$(\bfx)$ denote the greatest common divisor of the $x_i$. For integers $n,q\ge 2$, let ${\cal P}_{q,n}$ denote the set of all $q$-ary sequences $\bfx$ of length $n$ satisfying the following
properties:
\begin{enumerate}
\item $m(\bfx)=0$;
\item $M(\bfx)>0$;
\item GCD$(\bfx)=1$.
\end{enumerate}
\begin{theorem} For any $n,q\ge 2$, ${\cal P}_{q,n}$ {\it is an optimal Pearson code.} \end{theorem}
\beginofproof We will first show that ${\cal P}_{q,n}$ is a Pearson code. Property B is satisfied since any word in ${\cal P}_{q,n}$ contains at least one `0' and at least one symbol unequal to `0'. It can be shown that Property A holds by supposing that $\bfx \in {\cal P}_{q,n}$ and $\hat{\bfx} =c_1+c_2 \bfx \in {\cal P}_{q,n}$ for some $c_1,c_2\in \mathbb{R}$ with $c_2>0$. Clearly $c_1=0$, since  $c_1\neq0$ implies that $m(\hat\bfx)\neq 0$. Then, since $\hat{\bfx} = c_2\bfx$, we infer that GCD$(\hat{\bfx}) = c_2\times$GCD$(\bfx) = c_2$. Since, by definition, GCD$(\hat{\bfx}) = 1$, we have $c_2=1$ and conclude $\hat{\bfx}=\bfx$, which proves that also Property A is satisfied. We conclude ${\cal P}_{q,n}$ is a Pearson code.

We will now show that ${\cal P}_{q,n}$ is the greatest among all Pearson codes. To that end, let $\cal S$ be any $q$-ary Pearson code of length $n$. We map all $\bfx \in {\cal S}$ to $\bfx-m(\bfx)$ and call the resulting code ${\cal S}'$. Then, we map all words $\bfx'$ in ${\cal S}'$ to $\bfx'/$GCD$(\bfx')$. Note that both mappings are injective and that all words in the resulting code ${\cal S}''$ satisfy Properties 1-3. Hence, ${\cal S}''$ of size $|{\cal S}|$ is a subset of ${\cal P}_{q,n}$, which proves that ${\cal P}_{q,n}$ is optimal.
\endofproof
From the definitions of $T$-constrained sets and ${\cal P}_{q,n}$ it follows that
\beq {\cal S}_{q,n}(0,1)\subseteq{\cal P}_{q,n}\subseteq {\cal S}_{q,n}(0)\label{eqT2PT1}. \eeq
In the following subsections, we will consider the cardinality and redundancy of ${\cal P}_{q,n}$, and compare these to the corresponding results for $T$-constrained codes.
\subsection{Cardinality}
In this subsection, we study the size $P_{q,n}$ of ${\cal P}_{q,n}$. From (\ref{eqT2PT1}) and the remark following (\ref{eqN2}), we have
\beq N_2(q,n)\le P_{q,n} \leq N_1(q,n). \label{eqT2PT1size} \eeq
From Property B we have the trivial upper bound
\beq P_{q,n} \leq q^n-q, \label{eq 13} \eeq
which is tight in case $q=2$ as indicated in Section~\ref{description}, i.e.,
\beq P_{2,n} = 2^n-2. \label{eqbin} \eeq
In order to present expressions for larger values of $q$, we first prove the following lemma. We define $P_{1,n}=0$.
\begin{lemma}
For any $n\ge 2$ and $q\ge 3$,
\beq  \sum_{\substack{ {i=2} \\ {i-1|q-1}} }^{q} (P_{i,n}-P_{i-1,n})=q^n-2(q-1)^n+(q-2)^n, \label{eqDqn} \eeq
where the summation is over all integers $i$ in the indicated range such that $i-1$ is a divisor of $q-1$.
\end{lemma}
\beginofproof For each $i$ such that $2\le i\le q$ and  $i-1$ is a divisor of $q-1$, we define ${\cal D}_{i,n}$ as the set of all $i$-ary sequences $\bfy$ of length $n$ satisfying $m(\bfy)=0$, $M(\bfy)=i-1$, and GCD$(\bfy)=1$. Let ${\cal D}$ denote the union of all these disjoint ${\cal D}_{i,n}$.

The mapping $\psi$ from ${\cal S}_{q,n}(0,q-1)$ to $\cal D$, defined by dividing $\bfx\in{\cal S}_{q,n}(0,q-1)$ by GCD$(\bfx)$, is a bijection. This follows by observing that, on one hand, $\psi(\bfx)$ is a unique member of ${\cal D}_{(q-1)/{\rm GCD}(\bfx)+1,n}$, while, on the other hand, any sequence in $\bfy\in{\cal D}_{i,n}$ is the image of $((q-1)/(i-1))\bfy\in{\cal S}_{q,n}(0,q-1)$ under $\psi$.

Finally, the lemma follows by observing that $|{\cal D}_{i,n}|=P_{i,n}-P_{i-1,n}$ and $|{\cal S}_{q,n}(0,q-1)|=N_2(q,n)=q^n-2(q-1)^n+(q-2)^n$.
\endofproof\
We thus have with (\ref{eqDqn}) a recursive expression for $P_{q,n}$. Starting from the result for $q=2$ in (\ref{eqbin}), we can find $P_{q,n}$ for any $n$ and $q$. Expressions for $2\le q\le 8$ of the size of optimal Pearson codes, $P_{q,n}$, are tabulated in Table~\ref{tab1}. The next theorem offers a closed formula for the size of optimal Pearson codes, $P_{q,n}$. We start with a definition.

For a positive integer $d$, the M\"obius function $\mu(d)$ is defined~\cite[Chapter~XVI]{Ha1} to be $0$ if $d$ is divisible by the square of a prime, otherwise $\mu(d)=(-1)^k$ where $k$ is the number of (distinct) prime divisors of $d$.

\begin{theorem}\label{thm:Pearson_size}
Let $n$ and $q$ be positive integers. Let $P_{q,n}$ be the cardinality of a $q$-ary Pearson code of length $n$. Then
\begin{equation}\label{eqn:Psize}P_{q,n}=\sum_{d=1}^{q-1}\mu(d)\left(\left(\left\lfloor\frac{q-1}{d}\right\rfloor+1\right)^n-\left\lfloor\frac{q-1}{d}\right\rfloor^n-1\right).
\end{equation}
\end{theorem}

We use the following well-known theorem (see~\cite[Section~16.5]{Ha1}, for example) in our proof of Theorem~\ref{thm:Pearson_size}.

\begin{theorem}\label{thm:inversion_theorem}
Let $F:\bR\rightarrow\bR$ and $G:\bR\rightarrow\bR$ be functions such that
\[ G(x)=\sum_{d=1}^{\lfloor x\rfloor}F(x/d) \]
for all positive $x$. Then
\begin{equation}\label{eqn:inversion}F(x)=\sum_{d=1}^{\lfloor x\rfloor}\mu(d)G(x/d).\end{equation}
\end{theorem}

\begin{proof}[Proof (of Theorem~\ref{thm:Pearson_size}).] For a non-negative real number $x$, define
\[ I_x=\{0,1,\ldots,\lfloor x\rfloor\}=\bZ\cap [0,x].\]
Let $V_x$ be the set of vectors of length $n$ with entries in $I_x$ and with at least one zero entry and at least one non-zero entry. Define $G(x)=|V_x|$. To determine $G(x)$, note that there are $|I_x|^n$ length $n$ vectors with entries in $I_x$, and we must exclude the all-zero vector and the $(|I_x|-1)^n$ vectors with no zero entries. Since $|I_x|=\lfloor x\rfloor +1$, we find that
\begin{equation} \label{eqn:G} G(x)=|I_x|^n-(|I_x|-1)^n-1=(\lfloor x\rfloor +1)^n-\lfloor x\rfloor^n-1. \end{equation}
For a positive integer $d$, let $V_{x,d}$ be the set of vectors $\bc\in V_x$ such that $\GCD(\bc)=d$. Since $\bc\not=\mathbf{0}$, we see that $1\leq \GCD(\bc)\leq \max_i\{c_i\}\leq \lfloor x\rfloor$ and so  $V_x$ can be written as the disjoint union
\[
V_x=\bigcup_{d=1}^{\lfloor x\rfloor} V_{x,d}.
\]
Moreover, $|V_{x,d}|=|V_{x/d,1}|$, since the map taking $\bc\in V_{x,d}$ to $(1/d)\bc\in V_{x/d,1}$ is a bijection.

Define $F(x)=|V_{x,1}|$, so $F(x)$ is the number of vectors $\bc\in V_x$ such that $\GCD(\bc)=1$. Now,
\[
G(x)=|V_x|=\sum_{d=1}^{\lfloor x\rfloor}|V_{x,d}|=\sum_{d=1}^{\lfloor x\rfloor}|V_{x/d,1}|=\sum_{d=1}^{\lfloor x\rfloor}F(x/d).
\]
So, by Theorem~\ref{thm:inversion_theorem}, we deduce that \eqref{eqn:inversion} holds. Theorem~\ref{thm:Pearson_size} now follows from the fact that $P_{q,n}=F(q-1)$, by combining \eqref{eqn:inversion} and \eqref{eqn:G}.\endofproof
\begin{table}\caption{Size of optimal Pearson codes, $P_{q,n}$, for $2\le q\le 8$.}
$$\begin{array}{c|c} \hline
q & P_{q,n}  \\ \hline
2 & 2^n-2  \\
3 & 3^n-2^{n+1}+1 \\
4 & 4^n-3^n-2^{n+1}+3 \\
5 & 5^n-4^n-3^n+2  \\
6 & 6^n-5^n-3^n-2^n+4  \\
7 & 7^n-6^n-4^n+2^n+1  \\
8 & 8^n-7^n-4^n+3 \\
\hline \end{array}$$ \label{tab1}\end{table}

After perusing Table~\ref{tab1}, it appears that for $q \ge 4$, $P_{q,n}$ is roughly $q^n-(q-1)^n$. An intuitive justification is that among the $q^n$ $q$-ary sequences of length $n$ there are $(q-1)^n$ sequences that do not contain 0, which is the most significant condition to avoid. All this is confirmed by the next corollary.
\begin{corollary}
For any positive integer $q$, we have that
\[ P_{q,n}=q^n-(q-1)^n+O(\lceil q/2\rceil^n) \]
as $n\rightarrow\infty$.
\end{corollary}
\beginofproof
The $d=1$ term in the sum on the right hand side of \eqref{eqn:Psize} is $q^n-(q-1)^n$, and the absolute values of remaining terms are each bounded by $\lceil q/2\rceil^n$, since
\[ \lfloor (q-1)/d\rfloor+1\leq\lfloor (q-1)/2\rfloor+1=\lceil q/2\rceil.\qedhere\]
\endofproof

As discussed above, the 2-constrained codes ${\cal S}_{q,n}(0,1)$ and ${\cal S}_{q,n}(0,q-1)$ are Pearson codes. Therefore, it is of interest to compare $P_{q,n}$ with the cardinality $N_2(q,n)$ of 2-constrained codes. For $q\le 3$, we simply have ${\cal S}_{q,n}(0,1)={\cal P}_{q,n}$, and thus $N_2(q,n)=P_{q,n}$. However, for $q\ge 4$, we infer from (\ref{eqN2}), i.e., $N_2(q,n) = q^n-2(q-1)^n+(q-2)^n$, and Corollary~1, i.e., $P_{q,n}=q^n-(q-1)^n+O(\lceil q/2\rceil^n)$ that $N_2(q,n)<P_{q,n}$, with a possible exception for very small values of $n$. For all $q\ge 2$,
\beq P_{q,2} = N_2(q,2) = 2 \eeq
and it is not hard to show that
\beq P_{q,3}=6\sum_{j=1}^{q-1}\phi(j), \eeq
where $\phi(j)$ is Euler's totient function that counts the totatives of~$j$, i.e., the positive integers less than or equal to~$j$ that are relatively prime to~$j$.

We have computed the cardinalities of $N_1(q,n)$, $N_2(q,n)$, and $P_{q,n}$ by invoking (\ref{eqN1}), (\ref{eqN2}), and the expressions in Table~\ref{tab1}. Table~\ref{tabelqn} lists the results of our computations for selected values of $q$ and $n$.
\begin{table} \begin{center} \caption{\protect\small $N_2(q,n)$,   $P_{q,n}$, and  $N_1(q,n)$ for selected values of $q$ and $n$.}
\vspace*{0.3cm} \label{tabelqn} \begin{tabular}{ll|rrr}
\hline $n$ & $q$ & $N_2(q,n)$ & $P_{q,n}$ & $N_1(q,n)$\\\hline
 4&     4 &    110&   146&   175\\  4&     5 &    194&   290&   369\\
 4&     6 &    302&   578&   671\\ 5&     4 &    570&   720&   781\\
 5&     5 &   1320&  1860&  2101\\ 5&     6 &   2550&  4380&  4651\\
 6&     4 &   2702&  3242&  3367\\ 6&     5 &   8162& 10802& 11529\\
 6&     6 &  19502& 30242& 31031\\ 7&     4 &  12138& 13944& 14197\\
 7&     5 &  47544& 59556& 61741\\ 7&     6 & 140070&199500&201811\\
     \hline \end{tabular} \end{center} \end{table}
\subsection{Redundancy}
As usual, the redundancy of a $q$-ary code $\cal C$ of length $n$ is defined by $n-\log_q |{\cal C}|.$ From (\ref{eqN1}), it follows that the redundancy of a $1$-constrained code is
\begin{align}
r_1&=n-\log_q(q^n-(q-1)^n) \nonumber \\
&=-\log_q\left(1-\left(\frac{q-1}{q}\right)^n\right) \nonumber \\
&\approx  \left. \left(\frac{q-1}{q}\right)^n\right/ \ln(q),
\end{align}
for $n$ sufficiently large, where the approximation follows from the well-known fact that $\ln(1+a)\approx a$ when $a$ is close to 0. Similarly, from (\ref{eqN2}) we infer the redundancy of a $2$-constrained code, namely
\begin{align}
r_2&=n-\log_q(q^n-2(q-1)^n+(q-2)^n) \nonumber \\
&=-\log_q\left(1-2\left(\frac{q-1}{ q}\right)^n+\left(\frac{q-2}{ q}\right)^n\right) \nonumber \\
&\approx  \left. \left(2\left(\frac{q-1}{q}\right)^n-\left(\frac{q-2}{ q}\right)^n\right) \right/ \ln(q)
\end{align}
for $n$ sufficiently large. Since the $2$-constrained code ${\cal S}_{q,n}(0,1)$ is optimal for $q=2,3$, the expression for $r_2$ gives the minimum redundancy for any binary or ternary Pearson code. From Corollary~1, it follows for $q \ge 4$ that the redundancy of optimal Pearson codes equals
\begin{align}
r_{\rm P}&=n-\log_q\left(q^n-(q-1)^n+O\left(\left(\frac{q+1}{ 2}\right)^n\right)\right) \nonumber \\
&=-\log_q\left(1-\left(\frac{q-1}{ q}\right)^n+O\left(\left(\frac{q+1}{ 2q}\right)^n\right)\right) \nonumber \\
&\approx  \left. \left(\left(\frac{q-1}{ q}\right)^n+O\left(\left(\frac{q+1}{ 2q}\right)^n\right)\right) \right/ \ln(q).
\end{align}
In conclusion, for sufficiently large $n$, we have
\begin{equation}
r_P=r_2 \approx 2r_1
\end{equation}
if $q=2,3$, while
\begin{equation}
r_P\approx r_1\approx r_2/2
\end{equation}
if $q\ge 4$.
Figure~\ref{figredpearson} shows, as an example, the redundancies $r_1$, $r_2$, and $r_P$ versus $n$ for $q=8$ (the quantity $r_P$ was computed using the expression listed in Table~\ref{tab1}). Note that the redundancy $r_2$ decreases while the redundancy of prior art balanced codes, $r_0$, see (\ref{r_0}), increases with increasing codeword length $n$. The curve $r_0$ versus $n$ was not plotted in Figure~\ref{figredpearson} as the redundancy of balanced codes is much higher than that of Pearson codes. For example, an evaluation of (\ref{r_0}) shows that the redundancy $r_0=2.79$ for $q=8$ and $n=10$, while $r_P=0.147$ for the same parameters. %
\begin{figure} \centerline{\includegraphics[width=8cm]{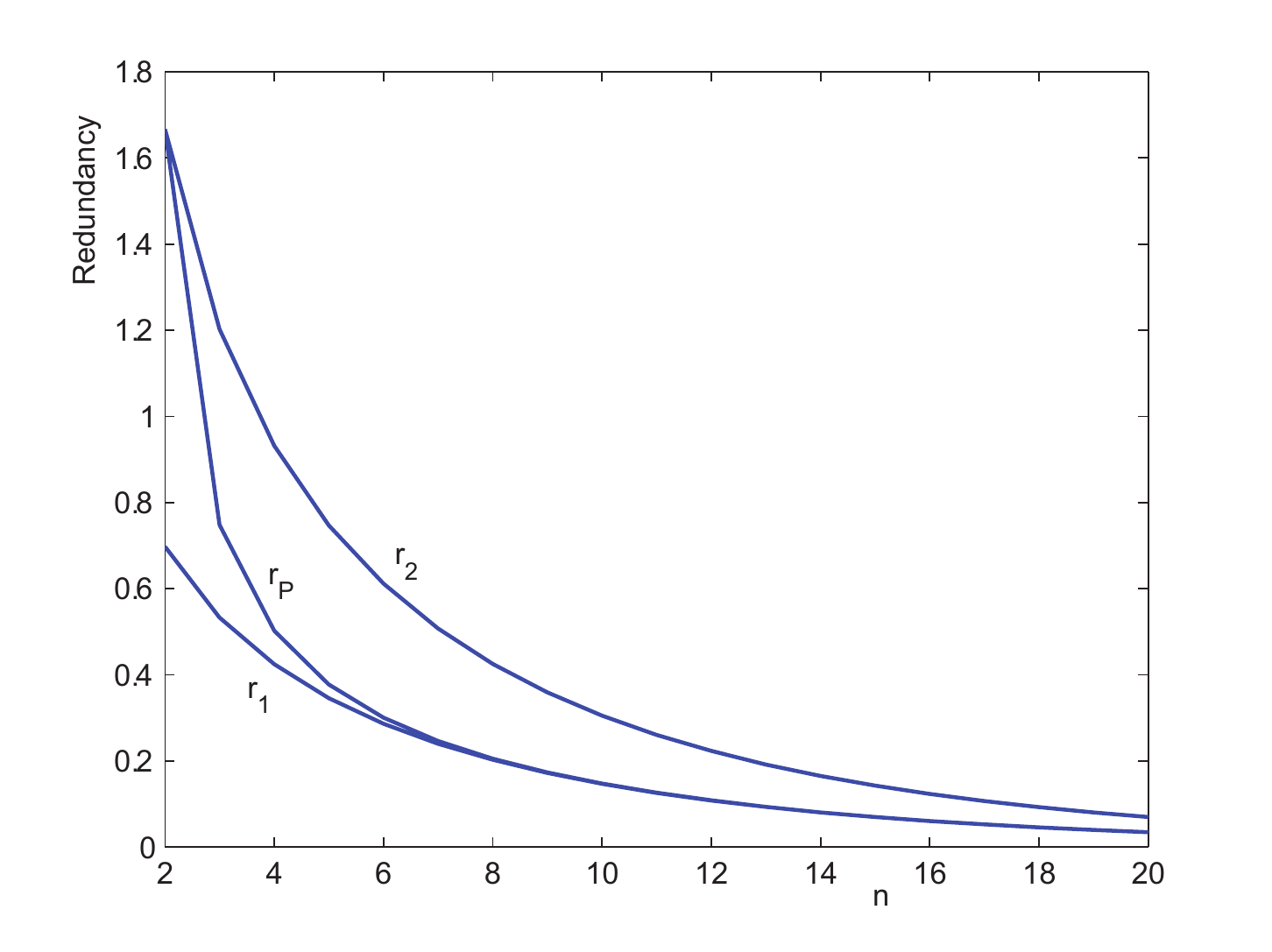}} \caption{\protect\small Redundancy $r_1$, $r_2$, and $r_P$ versus $n$ for $q=8$.}\label{figredpearson} \end{figure}
\section{Conclusions} \label{conclusions}
We have studied sets of $q$-ary codewords of length $n$, coined Pearson codes, that can be detected unambiguously by a detector based on the Pearson distance. We have formulated the properties of codewords in Pearson codes. We have presented constructions of optimal Pearson codes and evaluated their cardinalities and redundancies. We conclude that, except for small values of $q$ and/or $n$, the redundancy of optimal Pearson codes is almost the same as the redundancy of 1-constrained codes.

\vspace{1cm}

\noindent
{\bf Jos H. Weber} (S'87-M'90-SM'00) was born in Schiedam, The Netherlands, in 1961. He received the M.Sc. (in mathematics, with honors), Ph.D., and MBT (Master of  Business Telecommunications) degrees from Delft University of Technology, Delft, The Netherlands, in 1985, 1989, and 1996, respectively.

Since 1985 he has been with the Faculty of Electrical Engineering, Mathematics, and Computer Science of Delft University of Technology. Currently, he is an associate professor in the Department of Intelligent Systems. He is the  chairman of the WIC (Werkgemeenschap voor Inform\-atie- en Communicatietheorie in the Benelux) and the secretary of the IEEE Benelux Chapter on Information  Theory. He was a Visiting Researcher at the University of California at Davis, USA, the University of Johannesburg, South Africa, the Tokyo Institute  of Technology, Japan, and EPFL, Switzerland.  His main research interests are in the areas of channel and network coding, security, and (quantum) information theory.

\vspace{1cm}

\noindent
{\bf Kees Schouhamer Immink} (M'81-SM'86-F'90) received his PhD degree from the Eindhoven University of Technology. In 1998, he founded Turing Machines Inc, which has been successful in applying the tenets of information theory to digital data storage and transmission. He was from 1994 till 2014 an adjunct professor at the Institute for Experimental Mathematics, Essen-Duisburg University, Germany. In 1998, he founded Turing Machines Inc., an innovative start-up focused on novel signal processing for hard disk drives and solid-state (Flash) memories, where he currently holds the position of president.

Immink designed coding techniques of digital audio and video recording products such as Compact Disc, CD-ROM, CD-Video, Digital Audio Tape recorder, Digital Compact Cassette (DCC), Digital Versatile Disc (DVD), Video Disc Recorder, and Blu-ray Disc. He received a Knighthood in 2000, a personal Emmy award in 2004, the 1996 IEEE Masaru Ibuka Consumer Electronics Award, the 1998 IEEE Edison Medal, the 1999 AES Gold Medal, the 2004 SMPTE Progress Medal, the 2014 Eduard Rhein Prize for Technology , and the 2015 IET Faraday Medal. He received the Golden Jubilee Award for Technological Innovation by the IEEE Information Theory Society in 1998. He was named a fellow of the IEEE, AES, and SMPTE, and was inducted into the Consumer Electronics Hall of Fame, elected into the Royal Netherlands Academy of Sciences, the Royal Holland Society of Sciences, and the (US) National Academy of Engineering. He received an honorary doctorate from the University of Johannesburg in 2014. He served the profession as President of the Audio Engineering Society inc., New York, in 2003.

\vspace{1cm}

\noindent
{\bf Simon R. Blackburn} (M'12) was born in Beverley, Yorkshire, England in 1968. He received a BSc in Mathematics from Bristol in 1989, and a
DPhil in Mathematics from Oxford in 1992. He has worked in the Mathematics Department at Royal Holloway University of London since 1992,
and is currently a Professor of Pure Mathematics. His research interests include algebra, combinatorics and associated applications
in cryptography and communication theory.

\end{document}